\documentclass[prl,twocolumn,showpacs,preprintnumbers,amsmath,amssymb]{revtex4}
\usepackage{dcolumn}
\usepackage{bm}
\usepackage{epsfig}

\begin{document}
\title{Complex Quantum Phenomena in a Bilayered Calcium Ruthenate}
\author{G. Cao$^1$, L. Balicas$^2$, Y. Xin$^2$, E. Dagotto$^2$, J.E. Crow$^2$, C. S. Nelson$^3$, and J. P. Hill$^3$.}
\affiliation{$^1$Department of Physics and Astronomy, University of Kentucky, Lexington, KY 40506}
\affiliation{$^2$National High Magnetic Field Laboratory, Florida State University, Tallahassee, FL 32306.}
\affiliation{$^3$Department of Physics, Brookhaven National Laboratory, Upton, NY 11973.}
\date{\today}%
\begin{abstract}
Ca$_3$Ru$_2$O$_7$ undergoes an antiferromagnetic transition at $T_{\text{N}}=56 $K, followed by a
Mott-like (MI) transition at $T_{\text{MI}}=48$ K. This nonmetallic ground state, with a charge gap of 0.1 eV,
is suppressed by a highly anisotropic metamagnetic transition that leads to a fully spin-polarized metallic state.
We report the observation of Shubnikov-de Haas oscillations in the \textit{gapped} state, 
colossal magnetoresistance in the inter-plane resistivity with a large anisotropy different from that observed in 
the magnetization, and non-Fermi liquid behavior in the metallic state at high magnetic fields.
\end{abstract}
\maketitle

The colossal magnetoresistance (CMR) phenomenon, a gigantic decrease of resistance triggered by the application of a magnetic field,
so far has only been observed in manganese\cite{tokura} based oxides. Originally, this phenomenon was theoretically treated in terms of the
double-exchange (DE) mechanism\cite{basics}, whose basic description relies on the doping dependent valence 
states of Mn. However, to explain the richness of the experimentally observed phases, several instabilities, 
caused by a strong coupling between spin, charge and lattice degrees of freedom, have been proposed to compete with the ferromagnetic 
DE interaction. In fact, a series of experiments are explained\cite{elbio} assuming that the manganites are 
intrinsically \textit{inhomogenous} and present strong tendencies toward phase separation between ferromagnetic 
metallic (FMM) and antiferromagnetic charge/orbital ordered insulating (AFI) domains. 

This complex interplay between interactions, fluctuations, and inhomogeneities seems to be ubiquitous in transition
metal oxides (TMOs). Although their description is still the subject of an intense debate, it is clear that most of these systems cannot be described   
by the standard Fermi liquid (FL) theory with interactions renormalized in Landau like quasiparticles. Only recently a 
stoichiometric\cite{nigel} and an overdoped cuprate\cite{proust} have been 
claimed to present a FL-like ground state. In this respect, the 4d electron Ru-based stoichimetric oxides, or ruthenates, 
are perhaps unique systems since the oxygen vacancies inherent to the cuprates or the valence disorder peculiar to 
the manganites are \textit{not} characteristics of these systems. In fact, in at least one compound, 
Sr$_2$RuO$_4$\cite{maeno,mackenzie1} a FL ground state has been found. 
Despite the relative absence of disorder, the ruthenates are novel systems rich in complex physical phenomena as can be illustrated by the example of Ca$_3$Ru$_2$O$_7$\cite{cao,cao2,cao3}: With a 
double Ru-O layered orthorhombic structure\cite{cao}, Ca$_3$Ru$_2$O$_7$ (Ca327)belongs to the Ruddlesden-Popper series, 
Ca$_{n+1}$Ru$_n$O$_{2n+1}$, with $n=2$, where $n$ is the number of coupled Ru-O layers in a unit cell. Ca327 
undergoes an antiferromagnetic ordering at $T_{\text{N}}=56$ K followed by a sharp metal-nonmetal transition at 
$T_{\text{MI}}=48$ K\cite{cao2} (throughout this manuscript we define a ``non-metallic'' regime whenever the slope of the 
resivitity respect to the temperature $T$ is $d \rho/dT < 0$). The inter-layer spacing, c-axis, collapses at $T_{\text{MI}}$ \cite{nelson} 
suggesting a strong coupling between the electrons and the lattice. In between $T_{\text{N}}$ and $T_{\text{MI}}$ the system is in an antiferromagnetic 
metallic state (AFM)\cite{cao,cao2,cao3}, a rare occurrence for a stoichiometric compound at ambient pressure. 
At $T < T_{\text{MI}}$ Ca327 undergoes a first-order metamagnetic (MM) transition for a $B_{\text{MM}} \simeq 6$ T along
the a-axis, which is responsible for negative magnetoresistive. In this 
field-induced ferromagnetic phase the spins are nearly completely polarized\cite{cao,cao2}. 
At the MI transition, Raman spectra studies\cite{liu} reveal a rapid suppression of low-frequency electronic 
scattering closely associated with the formation of a charge gap $\Delta_c \sim 0.1 $ eV\cite{liu}. The corresponding gap 
ratio $R = \Delta_c /k_{B}T_{\text{MI}}\sim 23$ is large, suggesting that this transition is driven 
by strong electronic correlations, typical of a Mott-Hubbard system \cite{liu2,katsufuji}. 
In addition, the optical conductivity at room temperature yields a large scattering rate, 
$\hbar /2 \tau = 0.84$ eV, corresponding to a very short mean free path $\ell$ : $0.8\leq \ell \leq 8$ \AA. 
Remarkably, these values for $\ell$, assuming a typical Fermi velocity of $10^7-10^8$ cm/s \cite{puchkov}, are \textit{well beyond}
the so-called Mott-Ioffe-Regel limit for bandlike transport. 

In this letter, we present an electrical transport study of the Ca$_3$Ru$_2$O$_7$ compound at high magnetic fields ($B$). 
At low temperatures, very low frequency Shubnikov-de Haas oscillations are observed for $B$ nearly parallel to the 
c-axis \textit{only} in the ``insulating'' state. Consequently, Ca327 exhibits 
the \textit{coexistance} of quasiparticles with a ``gapped'' ground state. Below $T_{\text{MI}} = 48 K$, and for an angle dependent 
critical field in the b-c plane, a phase transition suppresses this gapped state stabilizing a ferromagnetic metallic phase. 
At the transition, colossal like negative magnetoresistance, assisted perhaps by a spin-valve type effect, 
is observed in the inter-layer resistivity $\rho_c$, what leads to a resistivity drop by a factor $\geq 10^3$. 
Thus, similarly to manganites, \textit{stoichiometric} Ru based oxide can also present CMR like effects, although
the underlying mechanism is perhaps unique.

Samples were grown in Pt crucibles by the self-flux method described in Ref.\cite{cao2}. 
The resistivity was measured by the standard Lock-In technique, while
the magnetization (susceptibility) was measured with a commercial SQUID magnetometer.  
Cryogenic facilities and magnetic fields up to 45 T were provided by the National High Magnetic Field Laboratory.

Fig. 1 (a) shows the inter-plane resistivity $\rho_{c}$ as a function of $B \|$c-axis for several temperatures $T$ between 0.6 and 6.5 K.
An oscillatory component, i.e., the Shubnikov-de Haas (SdH) effect, develops as $T$ is lowered. 
The inset shows the SdH oscillations at $T = 20$ mK in a limited field range $2 < B < 12$ T. Shown in Fig. 1 (b) is 
the amplitude of the SdH oscillations as a function of inverse field $B^{-1}$. Here, the SdH signal is defined as 
$(\sigma - \sigma_{b})/ \sigma_{b}$ where $\sigma$ is the conductivity ($1 / \rho_{c}$) and $\sigma_{b}$ is the 
background conductivity. The inset shows the amplitude of the SdH signal normalized respect to 
$T$ in a logarithmic scale, and as a function of $T$. 
\begin{figure}[htbp]
\begin{center}
\epsfig{file=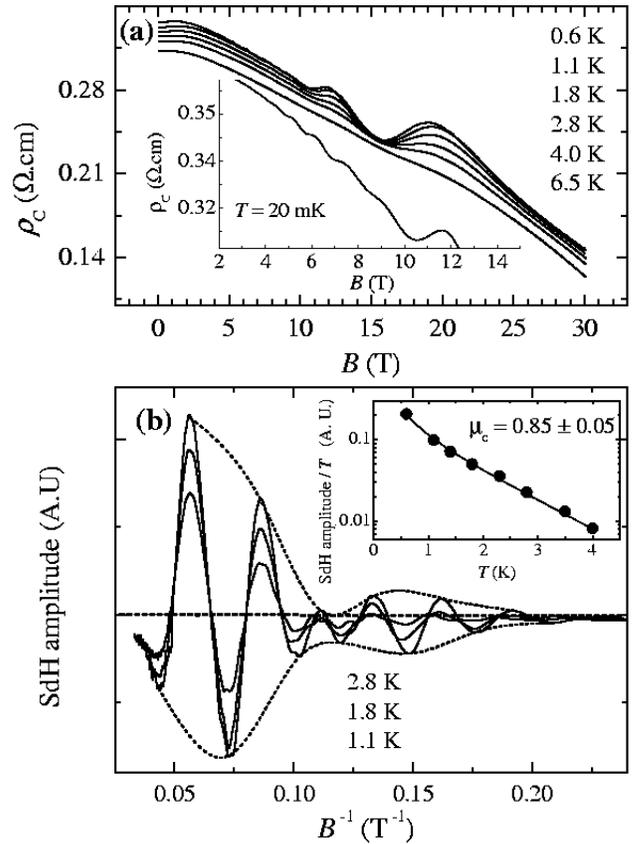, width=8.2cm}
\caption{(a) Inter-plane resistivity $\rho_c$ of a Ca$_3$Ru$_2$O$_7$ single crystal as a function of magnetic field $B$ and 
for several temperatures $T$. Inset: $\rho_c$ as a function of $B$ in a limited range and for $T=20$ mK. 
Notice the development of an oscillatory component as $T$ is lowered.
(b) The amplitude of the SdH oscillations as a function of $B^{-1}$ for several temperatures. 
Solid lines are a guide to eyes. Inset: SdH amplitude normalized respect to the temperature $T$ and as a function of $T$.
Solid line is a fit to LK formulae.}
\end{center}
\end{figure}
The solid line is a fit to the Lifshitz-Kosevich (LK) formulae, $x / sinh x$, where $x = 14.69 \mu_{c} T / B$. 
The fit is excellent and yields a cyclotron effective mass $\mu_{c}=0.85 \pm 0.05$. As clearly seen, the observed 
frequencies are very low, i.e., only a few oscillations are observed in the entire field range. 
A frequency of $F = 28$ T is determined by directly measuring the period of the oscillations for 
$B^{-1} \leq 0.1$ T$^{-1}$ which, based on the crystallographic data of Ca327\cite{cao,cao2} and on the Onsager relation 
$F_0 = A ( \hbar / 2 \pi e)$ ($\hbar$ is the Planck constant $e$ is the electron charge), corresponds to an area of  
$0.2 \%$ of its first Brillouin zone (FBZ). The vanishing oscillations at higher 
$B$ ($>$ 25 T) may indicate the proximity to the quantum limit which therefore imposes limitations on the applicability of 
the LK formalism \cite{lifshitz}. Dotted lines in Fig. 1 (b) are a guide to the eye showing the envelope of the 
SdH oscillations. Its shape suggests beating between two close frequencies, i.e., the presence of a second 
frequency $F_2 \sim 10$ T. 

QOs are usually observed in high purity compounds since a weak scattering of electrons preserves the sharpness of 
the Landau levels (LL). In TMOs the absence of QOs in the vast majority of compounds has been taken as an additional
evidence for a non-FL like, metallic state. At first glance the observation of QOs in Ca327 would seem impossible since 
it requires the existence of a robust FS and a \textit{long} $\ell$($ > 10^3$ \AA). In fact, from the QOs we can estimate 
the scattering rate $\tau^{-1}$ \cite{dingle} and the Fermi wave vector $k_F$, and from them we obtain $k_F \ell \gg 1$.
The fairly small $F\text{s}$ detected here formally excludes the possibility of QO's resulting from any source contamination. 
Furthermore, and as shown below, QO's \textit{are not} observed in the FMM state even in fields up to 45 T indicating 
that they are a property of the gapped state. Perhaps they can be attributed to small FS pockets resulting from FS 
reconstruction at $T_{\text{MI}}$, due, for instance, to an eventual spin/orbital ordering: A gap opens along most of the 
FS leaving (very) small-reconstructed electron and/or hole FS pockets. This would be consistent with the small value 
of $\mu_{c}$, i.e., a small density of states, indicating a small amount of charge carriers at the Fermi level. In fact, 
orbital ordering is predicted to occur in Ca$_2$RuO$_4$ \cite{hotta}, compatible with a sudden change in the basal plane at $T = 357$ K  
\cite{alexander,mizokawa}. But for Ca327, besides the collapse of the c-axis, no firm evidence 
for orbital ordering has been reported so far. At high fields the absence of QOs, or any other indication of FL behavior 
in the FMM state, clearly indicates the relevance of the electron correlations for this system. 

QOs are detected when $B$ is relatively close to the c-axis, but when $B$ is tilted towards the a or the b-axis 
we observe an orientation-dependent metamagnetic transition. For fields along the a-axis, a very sharp first order 
MM transition, observed at $B_{\text{MM}} \simeq 6$ T, see Fig. 2(a), destroys the antiferromagnetic order of the 
ferromagnetic Ru-O planes leading to a state with a magnetization $M_s = 1.7 \mu_B$/Ru along the a-axis\cite{cao2,cao4}.
For $B\|$ b-axis, the transitions are not only broader but also occur at higher $B$ than those for $B\|$ a (see Fig.2 (a)), 
implying a strong anisotropy between the a- and the b-axis which is consistent with the fact that the b-axis 
is the hard-axis for magnetization. Surprisingly, the decrease in $\rho_c$ for $B\|$ b is, 
at least, one order of magnitude larger than that for $B\|$ a. At low $T$, sample \#1 provides 
$\rho_c(0)/ \rho_c(30 \text{T}) \sim 10$ and $10^2$ for $B\|$a and $B\|$b, respectively. 
For $B\|$ a-axis, the effect of the MM transition on the inter-plane charge transport is considerably stronger than 
on the in-plane one, $ \rho_c (0) / \rho_c (H) > \rho_a (0) / \rho_a (H)$, which can be understood in terms of 
tunneling, or coherent motion, of (field-induced) spin-polarized electrons between Ru-O planes: 
Above $B_{\text{MM}}$, and due to the layered nature of Ca327, 
the spin-polarized (FM) Ru-O sheets, sandwiched between insulating (I) Ca-O planes, form a natural stack 
of FM/I/FM junctions that largely enhances the probability of inter-layer tunneling, and, thus, the 
inter-plane conductivity. Nevertheless, as seen in the figure, for fields along the b-axis, a much more pronounced resistivity drop 
is observed for $B \geq 14$ T. 
\begin{figure}[htbp]
\begin{center}
\epsfig{file=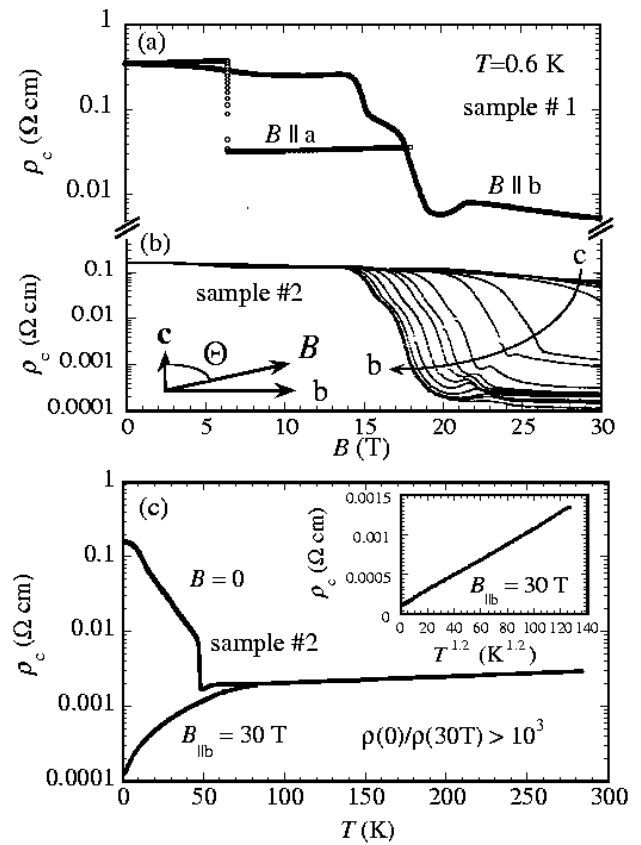, width=8.2cm}
\caption{(a) Inter-plane resistivity $\rho_c$ of a Ca$_3$Ru$_2$O$_7$ single crystal (sample \# 1) as 
a function of magnetic field $B$ along both the a and the b-axis at $T=0.6$ K. 
(b) $\rho_c$ from a second sample (sample \# 2) as a function of $B$ at $T=0.6$ K and for several angles $\theta$ in the b-c plane. 
The step in angle between traces is $\Delta \theta \simeq 7^\circ$. Lower panel: $\rho_c$ as a function of $T$ 
for two values of $B$, 0 and 30 T along the b-axis. Notice, how the gapped state is entirely suppressed by the field.
Inset: $\rho_c$ as a function of $T^{1.2}$ for $B = 30$ T along the b-axis. Notice the linear dependence.}
\end{center}
\end{figure}
In Fig. 2 (b) we show $\rho_c$ as a function of $B$ for a second sample (sample \#2) at $T=0.6$ K and for several 
angles $\theta$ between $B$ and the c-axis. Notice how $B_{\text{MM}}$ decreases as $B$ approaches 
the b-axis revealing the high degree of anisotropy of this transition. 
Thus, as indicated by Figs. 2 (a) and (b), the amplitude of the relative change in 
resistivity $ \rho_c (0) / \rho_c (H)$ is sample, $\theta$, and $T$ dependent, and can be as large as $10^3$. 
In fact, as seen in Fig. 2 (c), the CMR effect shown in Fig. 2 (b) is basically due to the \textit{complete} 
suppression of the gapped state by an external field (in our case, 30 T along the b-axis). 
This contrasts with what is observed for $B\|$ a, where a metallic state \textit{cannot} be stabilized at high fields:
$B$ completely polarizes the spins, broadens the transition towards the ground state, but has little effect on 
$T_{\text{MI}}$ which remains in the vicinity of 48 K. The application of $B \geq 14$T along b leads to the 
stabilization of a metallic state where, as seen in the inset, $\rho_c(T)$ shows an unusual, non-FL like, $T^{1.2}$ 
dependence on $T$. A sizeable amount of hysteresis is seen at $B_{\text{MM}}$ suggesting a first-order transition. 
By comparing the remarkable differences in behavior between both orientations, we conclude that spin-polarization 
alone \textit{cannot} account for the CMR like effects reported here.

The upper panel of Fig. 3 shows $\rho_c$ as a function of $B$ for several temperatures $0.6\leq T \leq 60$ K. The resistivity drop is strongly $T$ 
dependent: The higher $\rho_c (T)$, i.e., the lower the number of charge carriers thermally excited above the gap $\Delta_c$, the higher 
$\rho_c(0)/ \rho_c(H > 20)$. Notice how $B_{\text{MM}}$ decreases as $T$ increases. 

We summarize our results in the phase-diagram shown in 
the lower panel of Fig. 3. In this figure, perhaps the most relevant fact is the rather \textit{anisotropic} field-induced suppression of the charge
gapped state. Similar but isotropic behavior is observed in the RE$_{1/2}$Ca$_{1/2}$MnO$_3$ (RE= Pr, Nd, Sm,...) compounds\cite{kuwahara}, where the commensuration
of the band filling (doping level) with the crystal lattice spacing is believed to be responsible for their orbital/charge ordered ground state.
The application of a field produces a remarkable decrease in resistivity at a MM transition where a FM metallic phase
is stabilized at the expense of the charge/orbital ordered (OO) state. 
\begin{figure}[htbp]
\begin{center}
\epsfig{file=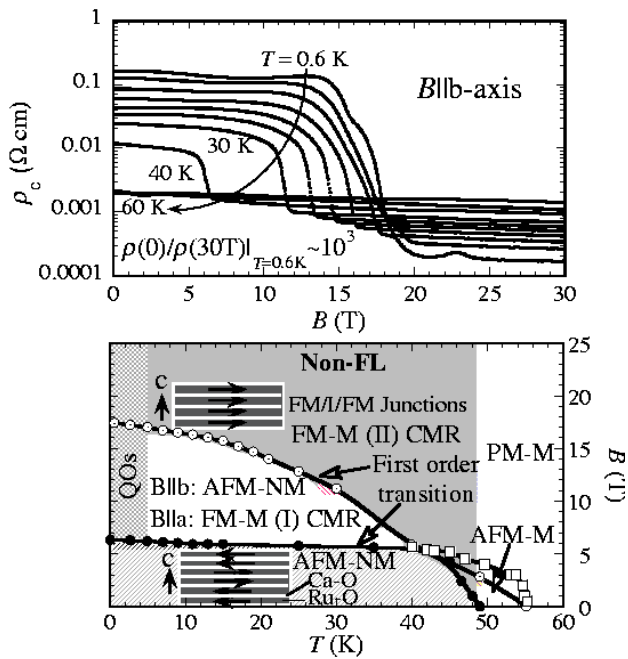, width=8.2cm}
\caption{Upper panel: $\rho_c$ as a function of $B$ parallel to the 
b-axis and for several temperatures $T$. The MM transition moves towards lower $B$ as $T$ increases.
Lower panel: The resulting $B-T$ phase diagram of Ca$_3$Ru$_2$O$_7$. 
At low fields, and as the $T$ is lowered, it shows two successive transitions, an 
antiferromagnetic metallic (AF-M) phase, followed by a gapped ground state, where the Ru-O planes seem to order in an 
antiferromagnetic A-type phase (AFM-NM). For $B\|$ a-axis, the AFM-NM phase is suppressed by a sharp MM transition at
 $B_{\text{MM}} \leq  6$ T (indicated by solid circles), which leads the system into a ferromagnetic poorly metallic phase (FM-M(I)).
For $B\|$ b-axis, the MM transition at $6 \leq B_{\text{MM}} \leq 15$ T completely ``melts'' the charge gap associated to the AFM-NM phase 
(opened circles), leading to a ferromagnetic metallic phase (FM-M(II)). The CMR effect is seen 
at the MM transition. Although the spins are nearly completely polarized along $B$, the FM-M(II) phase still shows an anomalously high 
resistivity and a nearly linear dependence on $T$ characteristic of a non-FL (gray area). At low $T$ and for $B\|$ b-axis, QOs are 
observed \textit{only} in the AFM-NM phase. At higher $T$, open squares show the frontier between the antiferromagnetic metallic (AFM-M) phase 
and, both, the paramagnetic metallic (PM-M) and the FM-M(II) phases.}
\end{center}
\end{figure}
From a thermodynamic point of view, this field-induced ``melting'' of the OO phase is understood by assuming that both states are 
\textit{almost} degenerate in energy, but under an external field $H$ the free energy of the FM metallic state may become lower due to the Zeeman 
term $-M_s H$ ($M_s$ is the spontaneous magnetization) inducing the hysteretic structural transition seen experimentally\cite{tokura}. 
It is interesting to mention that OO in the $t_{2g}$ orbitals is predicted to occur in ruthenates\cite{hotta}, 
and it has in fact, been reported in Ca$_2$RuO$_4$\cite{mizokawa}. In this framework, the gapped state of Ca327
could perhaps be the result of an OO transition at $T_{\text{MI}}$ while the CMR effect
would result from an eventual field-induced OO melting transition. This intriguing hypothesis certainly deserves further studies. 
Similarly to the bilayered Ca327, anisotropic CMR\cite{apostu} and spin-valve like effecs \cite{kimura,perring} have already been reported in 
bilayered manganites. Although we unveil here similarities between both families, the QOs and the highly 
anisotropic, doping independent, CMR like effect seen in Ca327 suggest that the physics of the ruthenates is indeed unique.
Alternatively, their study is a promissing new route for exploring the physics of CMR systems.

In summary, we have shown clear evidence for Landau quasiparticles coexisting with a gapped ground state in a transition metal oxide.
Furthermore, our study reveals that Ru based oxides can also display colossal magnetoresistance like behavior. 

We would like to thank the Hybrid Magnet Group at the NHMFL for their
invaluable assistance and V. Dobrosavljevic for very helpful discussions. 
The NHMFL is supported by a cooperative agreement between the
State of Florida and the NSF through NSF-DMR-0084173.
E. D. is supported by NSF grant DMR-0122523.



\begin{thebibliography}{}
\bibitem{dingle} From the amplitude of the QOs as a function of $B^{-1}$ one can extract,
the so-called Dingle temperature, $T_D= \hbar /2 \pi  k_{B} \tau$, a measure of the the relaxation time $\tau$ 
($k_{B}$ is Boltzmann constant). 
For our Ca$_3$Ru$_2$O$_7$ single crystals, $T_D$ ranges from 1.5 and 3.0 K, indicating their extreme high quality.
\end{thebibliography}

\end{document}